\documentclass[twocolumn,letter]{jpsj2}
\def\v#1{\mib #1}
\def\eff{\rm eff}

\newcommand{\ket}[1]{\left\vert {#1} \right\rangle}

\newcommand{\aver}[1]{\left\langle {#1} \right\rangle}

\def\runauthor{ Kazuo {\sc Hida} and Wei {\sc Chen}}

\title
{
Emergence of Long Period Antiferromagnetic Orders from Haldane Phase in $S=1$ Heisenberg Chains with $D$-Modulation}
\author
{
Kazuo {\sc Hida}\thanks{E-mail: hida@phy.saitama-u.ac.jp} and Wei {\sc Chen}$^1$
}

\inst
{Department of Physics, Saitama University, Saitama, Saitama, 338-8570
 \\
$^1$Institute of Biomedical Engineering,
University of Montreal,
Montreal, Qc. H3C 3J7
Canada
}

\recdate
{
\today
}

\abst
{
The effect of  spatial modulation of the single-site anisotropy $D$ on the ground state of the $S=1$ Heisenberg chains is investigated. In the case of period 2 modulation, it is found that the phase diagram contains the Haldane phase, large-$D$ phase, N\'{e}el phase of udud-type and u0d0-type. It is shown that the hidden antiferromagnetic order in the Haldane phase compatible with the spatial modulation of $D$-term get frozen resulting in the emergence of various types of N\'eel orders. The investigation of the model with longer period $D$-modulation also confirms this picture. }

\kword
{
S=1 Heisenberg chain, single-site anisotropy, spatial modulation,  Haldane phase, large-$D$ phase, N\'eel phase, hidden antiferromagnetic order
}

\begin{document}
\sloppy
\maketitle

\section{Introduction}

Antiferromagnetic quantum spin chains have been continually the subject of experimental and theoretical research in the past two decades. Among them, the Haldane ground state of the $S=1$ antiferromagnetic Heisenberg chain with a gap to the first excited state\cite{fd} has been extensively studied by many authors. This state is characterized by the hidden antiferromagnetic string order accompanied by the $Z_2\times Z_2$ symmetry breakdown. The easy plane single-site anisotropy $D (>0)$ destroys the Haldane ground state leading to the large-$D$ state without specific order. On the other hand, the easy axis  single-site anisotropy ($D <0$) drives the Haldane state into the N\'eel state.\cite{md,ht,chen} 

In this context, it is an interesting issue to investigate how the Haldane phase is modified if the  easy-axis and easy-plane $D$-terms coexist in a single chain.  As a simplest example of such competition, we first investigate the ground state phase diagram of the $S=1$ antiferromagnetic Heisenberg chain with alternating $D$-term,
\begin{eqnarray}
\label{ham0}
{\cal H} &=& \sum_{l=1}^{N}J\v{S}_{l}\v{S}_{l+1}+D_+\sum_{l=1}^{N/2}S_{2l-1}^{z2}\nonumber\\
&+&D_-\sum_{l=1}^{N/2}S_{2l}^{z2}, \ \ (J > 0)
\end{eqnarray}
where $D_+=D_0+\delta D, D_-=D_0-\delta D$ and $\v {S_{i}}$ is the spin-1 operator on the $i$-th site. The parameters $D_0$ and $\delta D$ represent uniform and alternating components of uniaxial single-ion anisotropy, respectively. The periodic boundary condition is assumed unless specifically mentioned. 

The most remarkable conclusion on the ground state of this model obtained in the present work is that the u0d0-type long period N\'eel state emerges for large $\delta D$ although this type of order is {\it not} compatible with the short range order in the Haldane phase at $D_+=D_-=0$. Here u, 0 and d stand for the single-site states $\ket{S^z_i}$ with $S^z_i=1, 0$ and $-1$, respectively. This implies that the {\it single-site} anisotropy can reverse the {\it sign} of spin-spin correlation in the corresponding isotropic model. This phenomenon can be understood as a selection of the specific antiferromagnetic order compatible with the $D$-alternation among the components of the Haldane state which is the superposition of various states with the hidden antiferromagnetic order. We also confirm this idea by investigating the similar model with long period $D$-modulation. 

This paper is organized as follows. In the next section, the effective Hamiltonians in various limiting cases are derived and the overall feature of the phase diagram is explained. The numerical exact diagonalization results and the quantitative phase diagram are presented in \S 3 . The model with long period $D$-modulation is discussed in \S 4. In the final section, we summarize our results.

\section{Effective Hamiltonians} 

Let us first consider the various limiting cases of the Hamiltonian (\ref{ham0}) in which the ground states can be determined by the effective Hamiltonians derived perturbationally. In the simplest limiting case of $D_+, D_- >> J$, all spins are confined to the state $\ket{0}$ and the ground state is obviously the large-$D$ phase. 

For $D_+, D_- << -J$, all spins are allowed to be in the states $\ket{\pm 1}$. The effective coupling between these spins are given by 
\begin{equation}
\label{hamudud}
{\cal H}^{(1)}_{\eff} = J\Delta\sum_{l=1}^{N}{S}_{l}^{z} {S}_{l+1}^{z},\ \ ({S}_{l}^{z}=\pm 1)
\end{equation}
up to the lowest order in $J$. Therefore we expect the udud-type N\'eel phase. 

In the limit of  $D_+ >> J, D_- << -J$, the spins on the odd-th sites are confined into the state $\ket{0}$ and those on the even-th sites can take $\ket{\pm 1}$ states. The latter spins are coupled via the Ising type {\it antiferromagnetic} coupling within the third order perturbation in $J$ as,
\begin{equation}
\label{hamu0d0}
{\cal H}^{(2)}_{\eff} = J_{\rm eff}\sum_{i=1}^{N/2}{S}_{2i}^{z} {S}_{2(i+1)}^{z}, \ \ J_{\rm eff}=\frac{J^3}{2\delta D^2} ,\ \ ({S}_{l}^{z}=\pm 1)
\end{equation}
although the spins $\v{S}_{2i}$ and ${S}_{2(i+1)}$ are {\it ferromagnetically} correlated in the absence of $D$-terms. The origin of the antiferromagnetic interaction is as follows. Let us denote the spin state of 3 successive sites of the original Hamiltonian (\ref{ham0}) by $\ket{S^z_{2i}\ S^z_{2i+1}\ S^z_{2(i+1)}}$. In the absence of the exchange term $J$, the ground states are $\ket{\pm1\ 0\ \pm 1}$ which are 4-fold degenerate. Due to the spin flip term in (\ref{ham0}), the $\ket{1\ 0\ 1}$ state is mixed with the $\ket{1\ 1\ 0}$ and  $\ket{0\ 1\ 1}$ states while the $\ket{1\ 0\ {-1}}$ state is mixed with the $\ket{1\  {-1}\ 0}$ and  $\ket{0\ 1\ {-1}}$ states. Therefore the energy of the $\ket{1\ 0\ 1}$ state is raised by the nearest neighbour pair of up spins in the intermediate state while the energy of the $\ket{1\ 0\ {-1}}$ state is lowered by the  nearest neighbour pair of up and down spins in the intermediate state. This means that the effective interaction between $S^z_{2i}$ and $S^z_{2(i+1)}$ is antiferromagnetic. It should be noted that the spin flip terms in the effective Hamiltonian appear only in the 4-th order perturbation in $J$. Therefore we expect the u0d0-type N\'eel phase in this regime. 

This can be also interpreted using the concept of the hidden antiferromagnetic string order in the following way. In the absence of the $D$-terms, the ground state has a hidden string order which implies that the spins with $\ket{\pm 1}$ are arranged antiferromagnetically if the sites with $\ket{0}$ are skipped.\cite{md,ht} The position of the sites with $\ket{\pm 1}$ and  $\ket{0}$ strongly fluctuate quantum mechanically and this antiferromagnetic order remains hidden because it is impossible to observe the correlation  between only the sites with  $\ket{\pm 1}$ experimentally. In the presence of strong $D$-terms, only the states consistent with the constraint set by the $D$-terms survive  among all states with hidden order. For $D_+, D_- << -J$, no  $\ket{0}$ sites are allowed, so that the hidden order is frozen into the explicit udud order. For $D_+>>J$ and $D_- << J$, the odd-th site must be in the state $\ket{0}$ and the even-th sites $\ket{\pm 1}$. To be compatible with the string order, spins must be arranged with u0d0 order which is no more hidden. Thus the strong $D$-term selects the spin states among the states with hidden order to realize the explicit N\'eel order of various periodicity.

\begin{figure}
\centerline{\includegraphics[width=70mm]{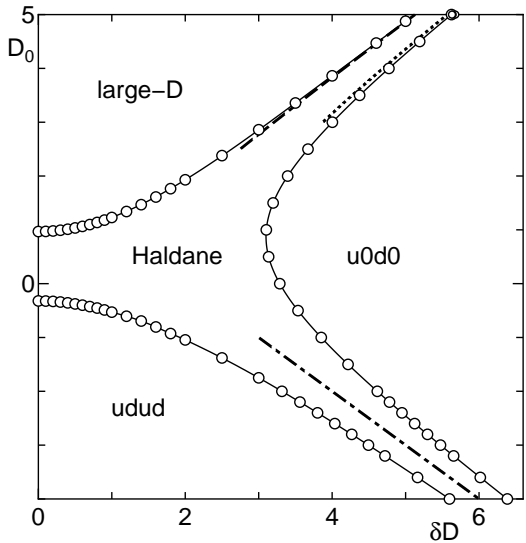}}
\caption{Ground state phase diagram of the Hamiltonian (\ref{ham0}). The error bars are within the size of the symbols. The solid lines are the guide for eye. The broken and dotted lines represent the approximate phase boundary given by Eqs. (\ref{hldapp}) and (\ref{hp4app}), respectively. The dash-dotted line is the line $\delta D = -D_0+2J$. }
\label{phase}
\end{figure}

Near the line $D_0 \sim \delta D >> J$, we can also derive the effective Hamiltonian. The spins on the odd-th sites are confined into the state $\ket{0}$ and those on the even-th sites can take all $\ket{\pm 1}$ and $\ket{0}$ states. The effective coupling between the latter spins are obtained within the second order perturbation in $J$ and first order in $D_0-\delta D$ as,
\begin{eqnarray}
\label{eff2}
{\cal H}^{(3)}_{\eff} &=& J_{\rm eff}\sum_{i=1}^{N/2}\Big\{\frac{1}{2}\left[{S}_{2i}^{+} {S}_{2(i+1)}^{-}+{S}_{2i}^{-} {S}_{2(i+1)}^{+}\right]\nonumber\\
&+&D_{\rm eff}{S}_{2(i+1)}^{z2}\Big\}
\end{eqnarray}
where
\begin{eqnarray}
J_{\rm eff}=\frac{2J^2}{D_0+\delta D} ,\ \  D_{\rm eff}=D_0-\delta D +\frac{2J^2}{D_0+\delta D}
\end{eqnarray}
For the model (\ref{eff2}), it is known that the large-$D$ phase is stable for $D_{\eff}/J_{\eff} > 0.35$. For $0.35 > D_{\eff}/J_{\eff} > -2.0$, the ground state is on the critical line between the Haldane and XY phase. For $D_{\eff}/J_{\eff} < -2.0$, the ground state is on the critical line between the N\'eel phase and the XY phase.\cite{md,ht,chen,kitaxy}  In the latter two cases, considering the continuity to other parts of the phase diagram, we may expect that the ground state is actually the Haldane phase or the N\'eel phase due to the Ising component of the effective exchange interaction which appear as higher order corrections in $J$. It should be noted that this N\'eel phase is the u0d0 phase in the original Hamiltonian (\ref{ham0}). The large-$D$-Haldane phase boundary is given by $D_{\eff}/J_{\eff} = 0.35$ which gives,
\begin{eqnarray}
D_0^2-\delta D^2 &=&-1.3J^2
\label{hldapp}
\end{eqnarray}
The u0d0-Haldane phase boundary is given by $D_{\eff}/J_{\eff} = -2.0$ which gives,
\begin{eqnarray}
D_0^2-\delta D^2 &=&-6.0J^2
\label{hp4app}
\end{eqnarray}
These lines are plotted by the broken and dotted lines in the ground state phase diagram in Fig. \ref{phase} along with the numerical results obtained in the next section. We thus expect that the Haldane phase shrinks but does not vanish even in the limit of large $D_0 \simeq \delta D$. 

On the other hand, for large negative $D_0$, the ground state is the almost perfect udud state for $D_0 < \delta D$ and the almost perfect u0d0 state for $D_0 > \delta D$. The ground state energy for the former case is given by
\begin{equation}
E_{\rm udud}=N(D_0-J)
\end{equation}
and in the latter phase
\begin{equation}
E_{\rm u0d0}=\frac{N}{2}(D_0-\delta D)
\end{equation}
Therefore the transition between these two phases should take place around $\delta D \simeq -D_0+2J$. Around this line, however, the  hidden antiferromagnetic order cannot freeze, because the correlation between the $2i$-th spin and $2i+2$-th spin can be ferromagnetic or antiferromagnetic depending on whether the $2i+1$-th spin is $\ket{0}$ or $\ket{\pm 1}$. Therefore the hidden order remains hidden. Accordingly a narrow Haldane phase should survive in the neighbourhood of the line  $\delta D \simeq -D_0+2J$ for arbitrarily large $\delta D$.

\section{Numerical Analysis}
\subsection{Haldane-Large-$D$ transition line }
This phase transition is a Gaussian transition. In order to determine the phase boundary  with high accuracy, we use the twisted boundary method of Kitazawa and Nomura.\cite{kita,kn} The Hamiltonian is numerically diagonalized to calculate the two low lying energy levels with the twisted boundary condition ($S^{x}_{N+1}=-S^{x}_1,$ $ S^{y}_{N+1}=-S^{y}_1,$ $ S^{z}_{N+1}=S^{z}_1$) for $N=8, 12$ and 16 by the Lanczos method.

For small $D_0$ and $\delta D$, the ground state is the Haldane phase with a valence bond solid (VBS) structure.\cite{aklt} Under the twisted boundary condition, the eigenvalue of  the spin reversal $T$ ($S_{i}^{z}\rightarrow -S_{i}^{z}, S_{i}^{\pm} \rightarrow -S_{i}^{\mp}$) is equal to $-1$ in this phase\cite{kita,kn}. As $D_0$ increases with positive value, a phase transition takes place from the Haldane to the large-$D$ phase for which  $T=1$.  Because the quantum number $T$ of Haldane and large-$D$ phases are different for twisted boundary condition,  the ground state energy of these two phases cross each other at the phase boundary.  For example, if $\delta D$ is fixed, the two levels cross at the finite size critical point $D_{0\rm c}(N)$.   Extrapolation to the thermodynamic limit is carried out assuming the formula $D_{0\rm c}(N)=D_{0\rm c}(\infty)+aN^{-2}+bN^{-4}$. 

\subsection{Haldane-udud and Haldane-u0d0 transition lines }

From symmetry consideration, these transitions are Ising type transitions. We therefore employ the phenomenological renormalization group (PRG) method to determine the transition lines. The Hamiltonian is numerically diagonalized to calculate the lowest energy gap $\Delta E(N, D_0, \delta D)$ with total $S^z=0$ in the periodic boundary condition using the Lanczos algorithm. Both N\'eel type states have $S^z=0$ and are two fold degenerate in the thermodynamic limit. For finite $N$, this degeneracy is lifted and the energy difference between them gives the smallest gap $\Delta E(N, D_0, \delta D)$ which decreases exponentially with $N$. On the other hand, in the Haldane  phase, the energy gap $\Delta E(N, D_0, \delta D)$ remains finite in the thermodynamic limit. Thus the product $N\Delta E(N, D_0, \delta D)$ increases (decreases) with $N$ in the Haldane(udud or u0d0) phase.   The intersection of $N\Delta E(N, D_0, \delta D)$ for two successive sizes $N=N_1$ and $N_2$ defines the finite size critical point $\delta D_{\rm c}(N_{1},N_{2})$  for fixed $D_0$. Taking into account the Ising universality class, these values are extrapolated using the formula $\delta D_{\rm c}(N_{1},N_{2})=\delta D_{\rm c}(\infty)+2C_{1}/(N_{1}+N_{2})^3$.\cite{mn,sakai} The same procedure can be also carried out interchanging the roles of $\delta D$ and $D_0$. For the Haldane-udud phase boundary, the system sizes $N=8, 10, 12, 14$ and 16 are used. For the Haldane-u0d0 phase boundary, only the sizes of multiples of 4 are compatible with the u0d0 order. We therefore use the data for $N=4, 8, 12$ and 16 to determine this phase boundary.

\begin{figure}
\centerline{\includegraphics[width=80mm]{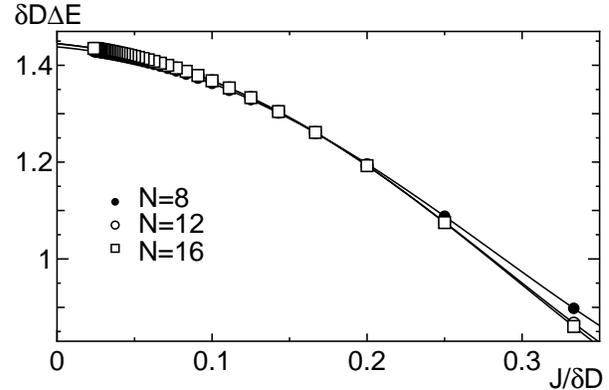}}
\caption{Energy gap $\Delta E$ on the line $\delta D = -D_0+2J$ plotted against $\delta D$. The vertical axis is multiplied by $\delta D$ to clarify the $\delta D$-dependence of $\Delta E$ The data for $N=12$ and $N=16$ overwrap within the size of the symbols. }
\label{halgap}
\end{figure}

The numerical phase diagram also shows that there exists an intermediate phase between the u0d0 phase and the udud phase which is continuously connected with the Haldane phase. This is consistent with the analytical argument in the preceding section. To further confirm the presence of this phase, we calculate the energy gap $\Delta E$ to the lowest excited state which has total $S^z=1$  along the line $\delta D = -D_0+2J$ up to $\delta D = 40J$ as shown in Fig. \ref{halgap} for $N=8, 12$ and 16. It is clear that the energy gap does not depend on the system size significantly and behaves as $\delta D^{-1}$ for large $\delta D$. We have also checked that the parity $T$ with respect to the spin inversion is always $-1$ with twisted boundary condition along the line $\delta D = -D_0+2J$. This confirms the VBS nature of this phase. 

\section{Long Period Modulation}

The mechanism of the selection of u0d0-phase given in \S 2 suggests that the similar mechanism also works for a variety of modulation patterns of $D$-terms producing a variety of long range N\'eel order. To confirm this idea, we also carried out the numerical calculation with the Hamiltonian with period $p$ modulation of the $D$-terms.
\begin{eqnarray}
\label{hamp}
{\cal H}_p &=& \sum_{l=1}^{N}J\v{S}_{l}\v{S}_{l+1}-\delta D\sum_{l=1}^{N/p}S_{pl}^{z2}\nonumber\\
&+&\delta D\sum_{l=1}^{N/p}\sum_{q=1}^{p-1}S_{pl+q}^{z2},
\end{eqnarray}
where $\v {S_{l}}$ are spin-1 operators. For $p=2$, this reduces to the Hamiltonian (\ref{ham0}) with $D_0=0$. The Haldane-N\'eel-type transition takes place at $\delta D_{\rm c} \simeq 2.69$ for $p=4$ (estimated from $N=8$ and 16), $\delta D_{\rm c} \simeq 2.75$ for $p=3$ (estimated from $N=6, 12$ and 18) and $\delta D_{\rm c} \simeq 3.29$ for $p=2$  (estimated from $N=4$, 8, 12 and 16). 

\begin{figure}
\centerline{\includegraphics[width=80mm]{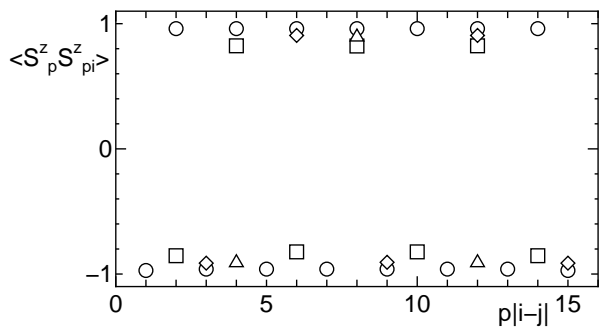}}
\caption{Correlation functions between the alive spins $\aver{S^z_{p}S^z_{pi}}$ for $p=1$($N=16$, circles), $p=2$($N=16$, squares), $p=3$($N=18$, diamonds) and $p=4$($N=16$, triangles). $\delta D=4$. }
\label{corr}
\end{figure}

For $\delta D > \delta D_{\rm c}$,  only the spins  $\v{S_{pl}}$ are alive. The spin-spin correlation functions among these {\it alive} spins are plotted in Fig. \ref{corr} for $\delta D =4$.  
It is clear that the alive spins are aligned antiferromagnetically with period $p$ irrespective of the short range correlation expected in the absence of $\delta D$-term. 

\section{Summary}

The ground state phase diagram of the spin-1 Heisenberg chains with alternating single-site anisotropy is determined by analyzing the  numerical diagonalization data and the perturbation analysis of various limiting cases. In addition to the Haldane, large-$D$ and conventional udud N\'eel phases, the N\'eel phase with u0d0 spin configuration is found. The emergence of these two types of N\'eel phases is explained as a selection of spin configuration compatible with the spatial modulation of $D$-terms among  various spin configurations with hidden antiferromagnetic string order in the Haldane phase. 

We find no direct transition between the udud and u0d0 phase and the intermediate gapful  phase is always present. This gapful phase is continuously connected with the Haldane phase. Between the large-$D$ phase and the u0d0 phase, we also find the Haldane phase. This robustness of the Haldane phase is also understood by the above picture.

The case of longer period $D$-modulation is also investigated to confirm the above picture. It is natural to expect various types of N\'eel order emerge from the Haldane state in the presence of various types of spatial modulation of $D$. 

So far, it is impossible to observe the string order itself experimentally, because the spatial spin structure corresponding to the string order is fluctuating quantum mechanically.  However, if the $S=1$ chain material with spatially modulated single-site anisotropy $D$ whose magnitude is larger than the exchange coupling $J$ is realized, the observation of the N\'eel order in such material can be regarded as the observation of the frozen string order.

The computation in this work has been done using the facilities of the Supercomputer Center, Institute for Solid State Physics, University of Tokyo, the Information Processing Center of Saitama University.  The diagonalization program is based on the TITPACK ver.2 coded by H. Nishimori.  This work is supported by a research grant from  a Grant-in-Aid for Scientific Research from the Ministry of Education, Science, Sports and Culture of Japan.


\begin{thebibliography}{99}
\bibitem{fd} F. D. M. Haldane: Phys. Lett. {\bf 93A} (1983); Phys. Rev. Lett. {\bf 50} (1983) 1153.
\bibitem{md} M. den Nijs and K. Rommelse: Phys. Rev. {\bf B40} (1989) 4709.
\bibitem{ht} H. Tasaki: Phys. Rev. Lett {\bf 66} (1991) 798.
\bibitem{chen} W. Chen, K. Hida and B. C. Sanctuary: Phys. Rev. {\bf B67} (2003) 104401.
\bibitem{kitaxy} A. Kitazawa, K. Hijii and K. Nomura : J. Phys. A: Math. Gen. 36 (2003) L351.
\bibitem{kita} A. Kitazawa : J. Phys. A Math. Gen. {\bf 30} (1997) L285.
\bibitem{kn} A. Kitazawa and K. Nomura: J. Phys. Soc. Jpn. {\bf 66} (1997) 3944.
\bibitem{aklt} I. Affleck, T. Kennedy, E. H. Lieb and H. Tasaki : Commn. Math. Phys. {\bf 115} (1988) 477; Phys. Rev. Lett. {\bf 59} 799 (1987).
\bibitem{mn} M. N. Barber: {\it Phase Transitions and Critical Phenomena 8}, ed. C. Domb and J. L. Lebowitz (Academic Press, London, New York, 1983) 146.
\bibitem{sakai} T. Sakai and M. Takahashi: J. Phys. Soc. Jpn. {\bf 59} (1990) 2688. 
\end{thebibliography}
\end{document}